# The Effect of Scale Consistency between Real and Virtual Spaces on Immersion in Exhibition Hybrid Spaces

Qiong Wu [1], Yan Dong [1*], Zipeng Zhang [1], Ruochen Hu [1],

1. *Department of Fine Arts,, Tsinghua University, Beijing, 100084, China*

**\* Corresponding author**,  dongyan0111@outlook.com

**Supported by**  the "Dual High" Project of Tsinghua Humanity Development (No.2023TSG08103)

## Abstract

In exhibition hybrid spaces, scale consistency between real and virtual spaces is crucial for user immersion. However, there is currently a lack of systematic research to determine appropriate virtual-to-real mapping ratios. This study developed an immersive interaction system based on Intel 3D Athlete Tracking body mapping technology. Two experiments investigated the impact of virtual space and virtual avatar scale on immersion. Experiment 1 investigated 30 participants' preferences for virtual space scale, while Experiment 2 tested the effect of 6 different virtual avatar sizes (25%–150%) on immersion. A 5-point Likert scale was used to assess immersion, followed by analysis of variance and Tukey HSD post-hoc tests.  Experiment 1 showed that participants preferred a virtual space ratio of 130% (mean 127.29%, SD 8.55%). Experiment 2 found that virtual avatar sizes within the 75%–100% range produced optimal immersion ($p < 0.05$). Immersion decreased significantly when virtual avatar sizes deviated from users' actual height (below 50% or above 125%). Participants were more sensitive to size changes in the 25%–75% range, while perception was weaker for changes in the 75%-100% range. Virtual environment slightly larger than real space (130%) and virtual avatars slightly smaller than users (75%–100%) optimize user immersion. These findings have been applied in the Intel Global Trade Center exhibition hall, demonstrating actionable insights for designing hybrid spaces that enhance immersion and coherence.

# 1 Introduction

As museum and exhibition space design increasingly embraces multi-dimensional experiences, hybrid spaces have become a significant focus of innovation. Recent advancements in Extended Reality (XR), motion capture devices, synchronized sensor data, and spatial mapping technologies have found widespread application in exhibitions, offering new immersive opportunities while introducing challenges such as offering new immersive opportunities while introducing challenges such as balancing virtual and real elements, as well as traditional and modern components. In practice, integrating virtual and real components often results in a disjointed experience. Familiar dimensions such as physical scale, movement patterns, and interaction styles shape users' cognitive models. Therefore, when virtual content diverges from these real-world expectations, it can trigger "cognitive dissonance," causing users to recognize the "virtual" nature of their experience. This recognition disrupts immersion and may lead to confusion in understanding the environment. Thus, ensuring seamless connectivity and consistency between virtual and real environments is essential to achieving genuine immersion.

The theory of hybrid spaces [1] posits that technological means can deeply integrate physical and media spaces, creating novel fields of interaction. A prime application of this theory is Spatial Augmented Reality (SAR), which embeds digital information into physical environments through screens or projections, enhancing immersive and interactive experiences in museum spaces [2]. However, such systems still face limitations related to physical constraints, struggling to balance enhanced immersion with natural interactions [3]. Projection systems are limited by line-of-sight issues, while wearable devices, though increasing interactivity, require additional user measurement, complicating interaction and often proving less suitable for exhibition contexts [4]. This paper, therefore, focuses on naturally embedding virtual content into real environments without reliance on wearable devices, aiming to create a seamless integration of virtual and real spaces that enhances immersion in exhibition hybrid spaces.

Designing for immersion in hybrid spaces involves several dimensions, including functionality [5, 6], interactivity [7-9], visual appeal [10, 11], spatial scale consistency [12-14]. Although these factors all impact user immersion, this paper focuses on the role of scale consistency between real and virtual spaces in enhancing immersion. Specifically, we examine how consistency in spatial proportions—between virtual elements and real-world references, as well as between users and their virtual avatars—optimizes spatial perception and the immersive experience for visitors in hybrid environments. In the design of hybrid exhibition spaces, spatial scale consistency is considered one of the key factors influencing immersion [15] because consistent scale enhances the naturalness of the transition between real and virtual, reducing to some extent the disruptions to immersion caused by cognitive dissonance [16]. Although the importance of spatial scale in blending real and virtual elements is recognized, there remains a gap in discussions on how to achieve consistent real and virtual scales in design practice.

This study further explores the balance of real and virtual spatial scale consistency to enhance immersion in hybrid exhibition spaces. Specifically, we focus on matching virtual and physical scales through coordinated hardware and spatial design to bring users' virtual experiences closer to the real world, thereby enhancing immersion without striving for a complete simulation of reality. The research employs a constructive design research approach, developing a body-mapping immersive hybrid space interaction system in an experimental setting to simulate a typical exhibition scenario. This scenario includes visual effects, interactive displays, and intelligent hardware applications to explore how immersive experiences in museum hybrid spaces can be supported and enhanced. Using Intel 3D Athlete Tracking (3DAT) technology, the system extracts key skeletal and posture data in real-time, allowing users to interact seamlessly with the virtual environment. Various experimental conditions assess the consistency between real and virtual scales to evaluate its impact on user immersion, with user feedback data collected for analysis.

Preliminary results indicate that aligning the scales of real and virtual spaces can enhance viewers' sense of immersion, allowing virtual elements to blend more naturally into the physical environment and reducing cognitive dissonance during the experience. Unlike traditional static displays, body-mapping technology captures users' movements in real-time, generating corresponding feedback within the virtual environment. This real-time interaction reduces perceptual divides between the virtual and real, creating a more cohesive hybrid space experience. Based on these findings, the system was showcased in Intel Global Trade Center (GTC) exhibition hall, demonstrating its potential to enhance immersive experiences and bridge virtual and physical spaces in real-world exhibition settings.

## 2 Related Research

Immersion refers to the system's ability to create a realistic illusion of reality within a virtual environment by engaging users' sensory and motor channels [7, 17]. This immersive experience enables users to feel as if they are interacting "naturally" with the virtual world, fostering a strong sense of "presence"—the impression of truly being "right there" in the virtual environment [10, 18]. In hybrid space research, consistency typically refers to the alignment between the virtual and real environments across various dimensions, including spatial scenes, objects, characters, interaction behaviors. These consistency dimensions are considered critical factors influencing immersion, as they are directly related to the user's spatial perception and interactive experience within the virtual world.

## 2.1 Spatial Scale Perception and Immersive Experience

Hybrid spaces integrate physical and virtual elements, and require design strategies that align with users' perceptual and cognitive processes to foster deeper immersive experiences. Spatial perception is central to users' understanding of spatial relationships, with scale perception serving as a crucial aspect. This perception determines users' awareness of relative sizes between themselves, surrounding objects, and the environment [13, 14]. Scale perception directly affects users' spatial understanding in virtual environments and thus plays an irreplaceable role in shaping immersion. Henry and Fummess [19] suggest that spatial perception involves three aspects: the size and shape of individual spaces, the observer's relative position within the overall layout, and the subjective experience of individual space. These components interact to influence perceptual consistency and the level of immersion users experience in hybrid spaces.

Factors such as physiological parameters (e.g., eye height and interpupillary distance), virtual object configuration, and the visual properties of the environment further impact spatial scale perception. For example, research by Dixon and Leyrer shows that lower eye height leads users to perceive objects as larger [20], which affects both egocentric distance perception and judgments of spatial scale [21]. By adjusting eye height and interpupillary distance, varying levels of spatial scale perception can be simulated, which in turn can alter user behavior [22]. In virtual environments, , which consist of geometric and lighting realism, also affects the user's subjective sense of presence [10]. Geometric realism refers to the accuracy of an object's shape and detail relative to the real world, while lighting realism refers to the fidelity of lighting and shadow effects [11].

Furthermore, Stevens et al. propose that users evaluate the consistency of environments not only through visual attributes but also through the experience of coexisting with virtual objects, which influences the perceived realism of the environment [23].Research also shows that even with spatial distortions, users maintain consistent size perceptions for familiar objects, which can serve as reference points for estimating the size of other objects [16]. In addition, Bhargava et al. indicate that the shape and size of virtual objects impact the accuracy of users' spatial scale judgments, with narrower objects leading to more accurate spatial estimates [12]. While existing studies have examined various factors that influence spatial scale perception, further research is needed to determine how these factors can be modulated to achieve consistent scaling between real and virtual spaces.

## 2.2  Perception of Virtual Avatar Scale Consistency and Immersive Experience

Virtual avatars are crucial for enhancing user immersion and interaction, with the design of their size and behavior having a particular impact on user's sense of body ownership and spatial consistency. Existing research has significantly investigated the size and behavior of virtual avatars in immersive spaces. Consistent avatar sizing allows users to establish a sense of ownership over their virtual avatar, thereby improving interaction accuracy. Anne Thaler et al. investigated users' visual perception of their avatars' dimensions and found that both male and female users estimated their avatar's size with relative accuracy within an error range of ±6%. Creating a proportionate virtual avatar can enhance users' physical self-awareness during immersive experiences [24]. Nina Döllinger et al. found that a virtual avatar fully matched to a user can transfer the users' internal feelings (perception) and self-empathy (cognition) to the avatar [9]. Sylvie Dijkstra-Soudarissanane used high-quality 3D motion capture and rendering techniques to control the size and actions of virtual avatars interacting with users, enabling elderly people individuals in immersive experiences to feel as though they were "together" with their family members represented as virtual avatars [24].

Consistent avatar actions provide users with physical comfort and help to avoid problems such as dizziness [25]. Kessler and Rutherford [26] suggest that spatial perspective taking is an embodied process that relies on action-related and proprioceptive representations, and its benefits become measurable when users are allowed to move within a scene to enhance their skills. Junjian Chen et al. used RGB webcams to capture players' natural movements and map them onto virtual characters within a game [27]. This approach ensured motion consistency between players and their avatars, reducing cognitive load and increasing immersion. In addition to manipulating users' viewpoints and objects, some studies have investigated the effects of changing the proportions of virtual body parts, such as hands, feet and self-avatars [28]. In particular, users of different sizes experience significant differences in perception. While the construction and perception of virtual avatar size in immersive spaces has been widely studied, limited attention has been paid to the effects of controlling avatar size and behavior in virtual environments on the immersive experience.

Although existing research has explored factors influencing spatial scale perception, such as spatial dimensions, virtual avatars and relative positioning, the design of exhibition hybrid spaces lacks systematic data-driven studies of real-virtual mapping ratios. Current approaches to setting scale ratios in virtual and mixed reality spaces largely rely on designers' experiential judgement, which often fails to ensure consistency and appropriateness, especially when considering different user characteristics and spatial perception needs. Thus, the establishment of real-to-virtual mapping ratios based on quantitative analysis, with parameterised studies providing scientifically based guidelines, remains an unresolved issue in the field. Such research can provide designers with more specific and reliable guidance, allowing for greater immersion and consistency in harmonising virtual and real scale ratios. Consequently, our work aims to address the following questions:

**RQ1:** How can we design adaptable virtual-to-real scale ranges to enhance users' perceptions of spatial consistency across different user groups?

**RQ2:** In virtual embodiment design, how can adjustments to the mapping ratio between virtual avatars and users optimise user immersion in hybrid spaces?

# 3 System Description

## 3.1 System Design

We propose an immersive hybrid space interaction system that uses body-mapping to enable authentic user engagement with the virtual environment (Figure 1). By aligning spatial scale, virtual avatar and interaction feedback, this system attempts to create the illusion of 'being present' in the virtual environment.

On the one hand, we modify spatial dimensions, visual anchors and other virtual design elements to enhance the spatial coherence between virtual and physical environments. On the other hand, we adapt body-mapping technology to allow the user to control the actions of a virtual avatar through physical movement, optimising the user's likeness and immersion in the avatar through spatial scale, relative positioning and interaction feedback to enhance the perception of engagement with both virtual and physical worlds.

The system task requires the user to navigate a constantly changing digital world, collecting virtual chips scattered throughout the environment to increase movement speed and avoid obstacles in order to reach an endpoint as quickly as possible. The core mechanism requires the user to constantly adjust their position and posture, and to remain in close interaction with the evolving virtual environment. As the user collects more chips, dynamic changes within the virtual scene - such as intensified visual effects, the appearance or disappearance of obstacles and variations in sound design - occur in synchrony, increasing the user's sense of impact on the virtual world.

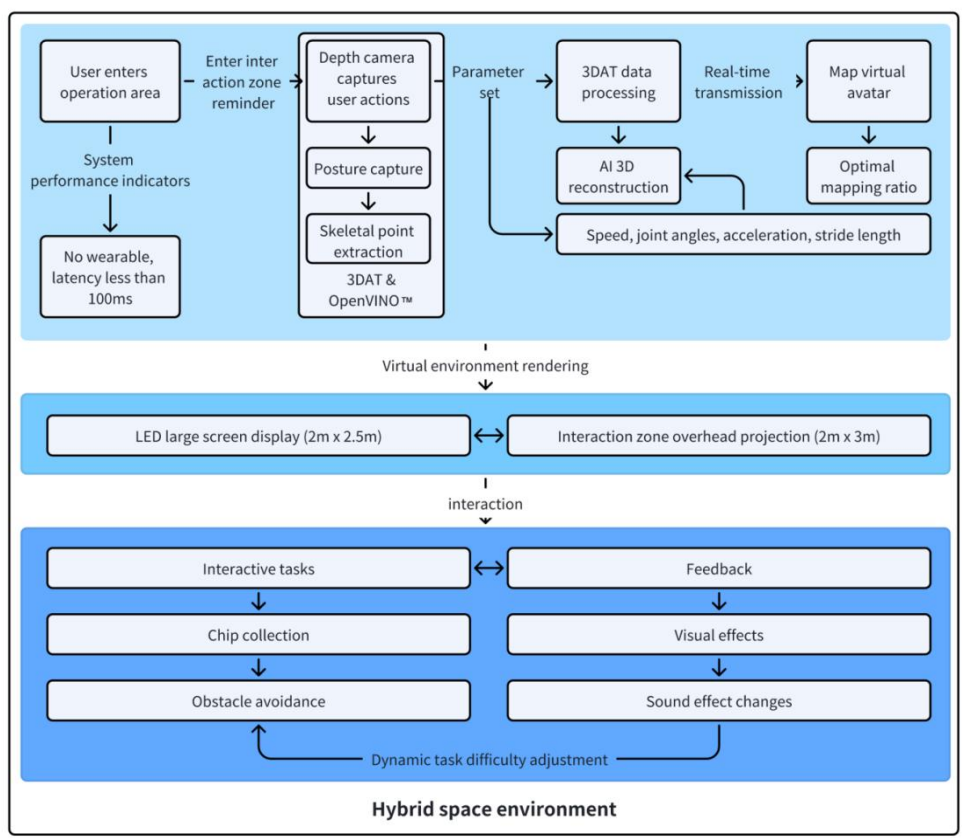

**Figure 1   Hybrid Space Interaction System Architecture.**

The system consists of three phases: **Information Extraction** (capturing visual and postural data), **Information Expression** (presenting information within the hybrid space), and **Interaction** (facilitating effective exchanges between the virtual and real environments). The functions of each stage are as follows:

### 3.2 Information Extraction

This phase is critical to achieving immersive interaction within the hybrid space. By combining Intel 3DAT technology with a depth camera, the system enables precise and efficient tracking of human movement. The depth camera, positioned approximately 3 meters from a large LED screen (2 m x 2.5 m), covers a 3 m x 2 m interactive area, allowing the user to move freely within this zone without any wearable devices. The system captures real-time posture data and records motion trajectories, which are processed via OpenVINO™ to extract key skeletal information. Unlike traditional Kinect systems, our solution integrates AI-driven 3D reconstruction technology to provide highly accurate analysis of motion parameters such as velocity, joint angles, acceleration, and step length. The extracted data are mapped in real-time to a pre-configured virtual avatar model, ensuring the virtual avatar accurately reflects the user's movements, with latency kept under 100 ms.

Body-mapping technology significantly enhances the user's sense of embodiment within the virtual avatar, effectively supporting the system's core task: encouraging users to adjust their position and movement to stay in close interaction with the dynamically evolving virtual environment. Leveraging OpenVINO™ hardware acceleration, the system performs complex AI computations on edge devices, enabling real-time dreading, pre-processing and analysis of data. This approach reduces system latency and improves responsiveness, providing a smooth interactive experience for users. The system maintains consistent performance under varying lighting conditions, making it adaptable to a range of indoor environments. It provides high-quality, low-latency input data that effectively supports interaction designs such as virtual chip collection and obstacle avoidance within the system tasks.

### 3.3 Information Expression

*3.3.1 Hybrid Space Environment* The system is based on a real exhibition space, combined with a 2 m x 2.5 m LED screen and a 3 m x 2 m ceiling projection to construct the virtual scene. Elements in the virtual environment are designed with reference to real-world visual cues, using dynamic light sources and material reflections to simulate real physical properties. Object shape and shadow changes help users perceive spatial depth and object distance, enhancing visual realism and reducing cognitive gaps between virtual and real environments. The virtual space creates a sense of infinite extension through linear perspective, extended lines, and motion, guiding users' visual focus and making the virtual scene a dynamic extension of the physical space. Carefully set spatial scales and referent object sizes further enhance the realism of the virtual environment, making the transition from real to virtual more natural for users while effectively reducing potential discomfort in the virtual environment.

*3.3.2 Virtual Avatar* To enhance immersion, the system's virtual avatar is carefully designed to simulate real-world scale as much as possible, allowing users to easily project themselves into the virtual avatar. The visual quality of lighting effects, interfaces, and materials in the virtual environment is are matched to the proportions of the virtual avatar to ensure the visual credibility of the scene. Moreover, the system enhances the user's embodied experience by accurately mapping the user's body movements to the virtual avatar. Based on action-related and proprioceptive representations, it fully exploits the benefits of spatial perspective taking to help users better understand the spatial layout and relative object positions [D10 23], thereby enhancing the user's sense of presence in the scene.

## 3.4 Interaction

The challenge in the interaction stage of this system is how to promote effective connection between virtual and real through interaction design in a hybrid space without relying on physical devices, establishing consistent perception. Firstly, the system not only ensures high consistency between users' physical movements and their virtual avatars, but more importantly creates an immersive illusion through alignment between user actions and corresponding feedback. Effects such as avoidance feedback and collision responses are conveyed through synchronised visuals and sound, giving users a heightened sense of interactivity and deepening their sense of presence within the scene during tasks.

Secondly, while the users move in parallel in a relatively fixed area of real space, the main direction of movement in the mapped virtual space is in depth. To reduce the sense of contradiction caused by directional inconsistency, the system uses the relative motion of the LED screen and the projection screen to create a sense of high-speed motion. It adjusts the relative speed for acceleration and deceleration based on the user's task performance, combined with dynamic visual cues (such as background blur, acceleration effects) to further enhance the user's sense of motion. On the other hand, floating platforms that wobble according to the user's position are added to the virtual space to alleviate directional contradictions due to the relative position of objects.

Furthermore, in terms of task design, the system dynamically adjusts the difficulty of the task in real time based on user's performance. For example, the number, speed, and frequency of obstacles change according to the user's current state to ensure the task remains challenging and interesting.

## 4 Experiment

### 4.1 Preference for spatial mapping scale

*4.1.1 Experimental Design* There is a lack of in-depth research on spatial mapping ratios, with no clear upper or lower limits established. To address this gap, this study systematically examines various spatial mapping ratios and their impact on user immersion and interaction accuracy. By testing different ratio configurations, we aim to identify an optimal range that closely aligns virtual experiences with real-world spatial perceptions. Therefore, we use the physical environment as a benchmark in our experiment. Participants adjust the parameters of the virtual environment based on their sense of immersion under the guidance, starting from a 1:1 ratio. The virtual environment displayed on the interactive screen is set to match the physical screen height of 2.5 metres.

*4.1.2 Hypotheses*

1. At a spatial mapping ratio of 1:1, users will experience the strongest immersion, and participants' preferred scaling ratios will cluster around 100%.

2. As the spatial mapping ratio increases, users' sense of spatial immersion may initially strengthen. However, excessively large ratios may produce a sense of distortion.

*4.1.3 Experimental Procedure* This experiment uses a within-subjects design, starting with the virtual space set to a 1:1 spatial mapping ratio and a focal length equivalent to a 50 mm lens. Participants stand at a designated position, 3 metres from the large screen. Following participants' instructions, trained operators adjust the size of the virtual space using a controller to achieve the ratio that each participant finds most immersive. In this context, immersion refers to the feeling of actually existing in the virtual environment and interacting with objects, rather than merely observing them. After each adaptation, there is a brief 5-second blackout, during which the experimenters record the final spatial ratio value and the time taken by participants to complete the adaptation (Figure 2). The only light source in the experimental environment is the large screen, which is set to 450 nits based on pre-test results, with constant brightness and colour temperature. All experiments were conducted in a laboratory with controlled lighting to prevent external light from interfering with immersion.

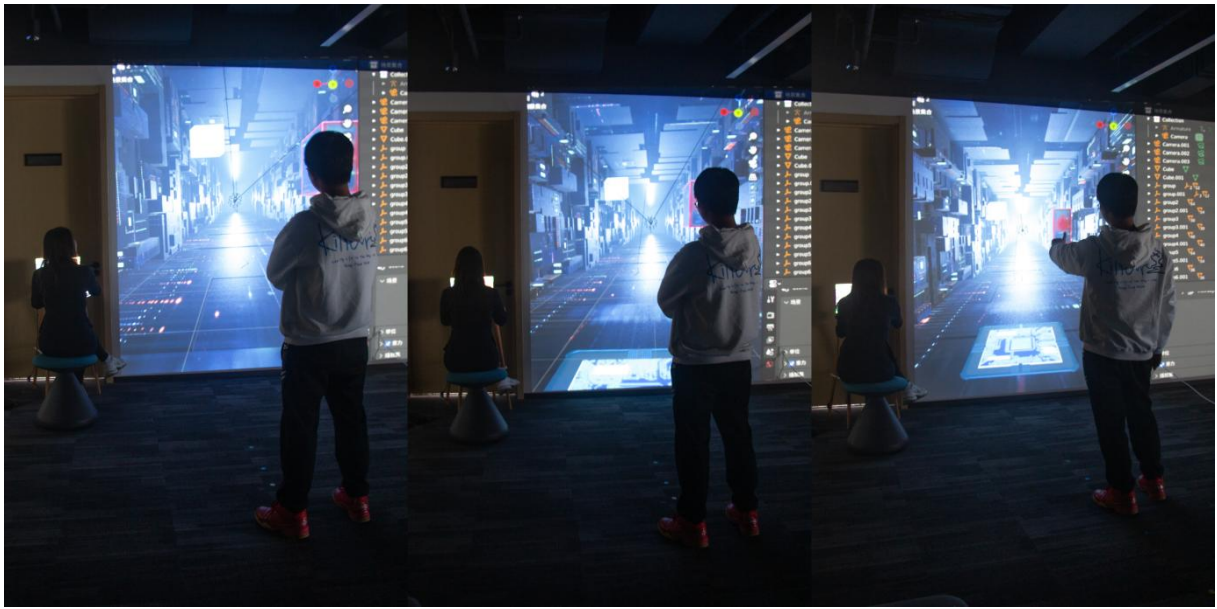

**Figure 2   Experimental Setup for Spatial Mapping Scale Preference**

*4.1.4 Participants* This study recruited 30 healthy adult participants, including 15 males and 15 females, aged 20 to 27 years (mean age 24 years, standard deviation 2.3 years). All participants were right-handed with normal or corrected-to-normal vision. To ensure experimental validity, participants were screened during recruitment to ensure that they would not experience dizziness, nausea, or other symptoms of motion sickness in virtual reality environments. All participants signed an informed consent forms before the experiment. The sample size of 30 participants was selected to ensure adequate statistical power. By allowing participants to directly adjust the virtual space scale, we randomly sampled from all possible spatial mapping ratios to collect subjective preference data from participants.

*4.1.5 Result* Based on the recorded data (Appendix A Table 1), we plotted the distribution of the average spatial mapping ratio (Figure 3). The green bars represent the actual distribution of the data. The width of each bar represents an interval (bin), while the height represents the density of data points within that interval. Here, density refers to the normalized frequency of observations in each bin, allowing the total area of all intervals to sum to 1, which helps to visually compare the relative concentration of different intervals. Compared to absolute frequencies, density better reflects the distribution pattern of the data.

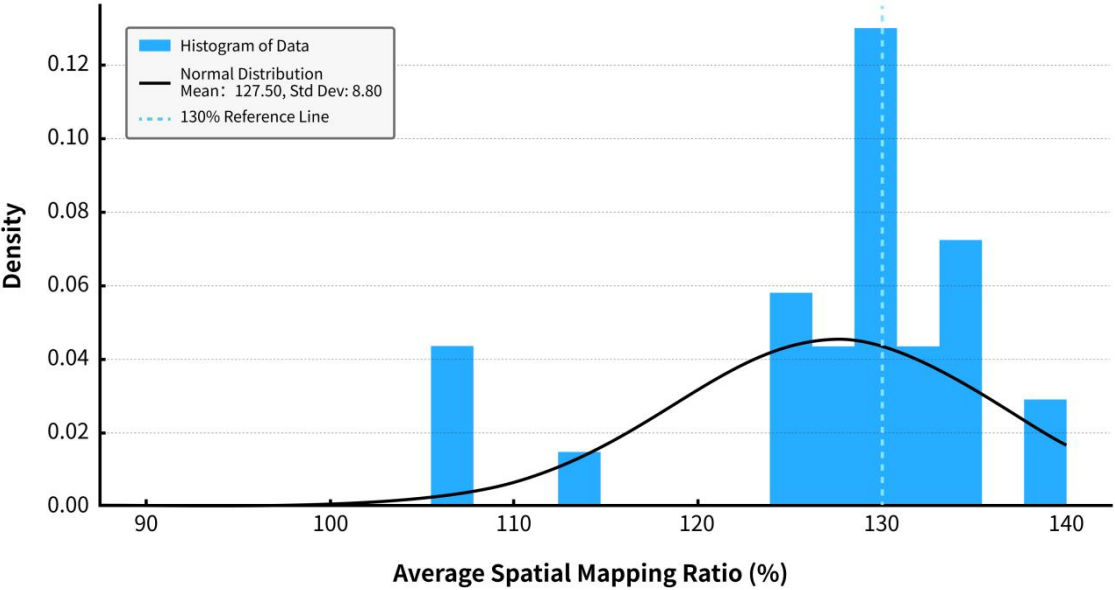

**Figure 3**   **Histogram and normal distribution fit for average spatial ratio**

Experimental data show that participants' choices when adjusting the spatial mapping ratio were concentrated around 130%. For the spatial scaling ratio, the mean value was 127.29%, with a standard deviation of 8.55%, a minimum value of 106.10%, and a maximum value of 138.33%. The small standard deviation indicates a strong central tendency in the data distribution. For the average adjustment time, the mean was 31.10 seconds, with a standard deviation of 5.09 seconds, a minimum of 21.98 seconds, and a maximum of 43.83 seconds. Furthermore, approximately 68.3% (20 out of 30 participants) of the data points fell within one standard deviation (i.e., between approximately 127% and 133%). This indicates that the spatial ratio around 130% is indeed the preferred concentration area for participants, corresponding to the range where most people felt a stronger sense of immersion.

Although we hypothesized that a 1:1 mapping ratio (i.e., 100%) would provide the most natural immersive experience as it directly simulates the real space, the experimental results show that participants preferred a ratio of around 130%. This might be because a slightly exaggerated spatial ratio enhanced the visual impact and sense of immersion in the environment. In the 135%-140% range, the frequency of participant choices decreased by about 40% compared to the 130% range, indicating that most participants reported a slight sense of distortion at larger ratios. Especially beyond 140%, the frequency of choices further decreased significantly, supporting Hypothesis 2 that as the ratio increases, the sense of immersion might increase to some extent, but beyond a certain threshold, a sense of distortion begins to emerge and affect the experience.

### 4.2  Preference for virtual avatar mapping size

*4.2.1 Experimental Design* In addition to spatial scale, the size of virtual avatars is also an important factor affecting immersion. To investigate immersion in virtual environments more comprehensively, we introduced virtual avatars in Experiment 2, building upon Experiment 1. These are 3D character models controlled by participants through camera motion capture devices. This experiment aims to explore the optimal range of virtual avatars size and study how different size affect user immersion. According to existing research, when users experience virtual worlds at small size (such as with 80cm or 30cm bodies), their perception of object size and distance changes [29]. Furthermore, studies have shown that when the height of a virtual avatar is 20% taller than the user's actual height, the user's ability to judge passage in dynamic environments is not significantly affected [30]. Based on these findings, we set the virtual avatar size range from 25% to 150% (relative to participants' real height) to cover a wide range of mapping size changes. In the experimental design, we used an evenly divided grid method, dividing sizes into 6 levels (25%, 50%, 75%, 100%, 125%, and 150%), covering extreme cases of reduction and enlargement. As this experiment is conducted in a large screen environment, unlike VR close-up viewing, we do not consider the effect of interpupillary distance on immersion for now. At the same time, we will record users' heights to analyze their potential impact on immersion.

*4.2.2 Hypotheses*

1. When the height of the virtual avatars is closer to the participant's actual height (e.g., between 75%-125%), users will experience stronger immersion.

2. As the size of the virtual avatar gradually deviates further from the user's actual height (below 50% or above 125%), user's sense of immersion will decrease significantly. At these extreme ratios, users may experience a sense of distortion in objects and spaces within the virtual environment, making their interactive experience less natural than at ratios close to 1:1.

*4.2.3 Experimental Procedure* In Experiment 1, we found that the optimal spatial mapping ratio was approximately 130%. Therefore, in this experiment, all spatial dimensions of the virtual environment are fixed at a 130% scaling ratio to ensure environmental consistency. Lighting conditions in the laboratory are kept constant (only from the large screen at 450 nits brightness) to ensure experimental repeatability and environmental stability. Participants' actual heights are measured and used for subsequent calculations of virtual avatar size.

The height of the virtual avatar is set sequentially to six different sizes: 25%, 50%, 75%, 100%, 125% and 150% of the participant's actual height (Figure 4). At each size, participants are required to complete three sets of tasks:

1. **Static perception**: Remain still in the virtual environment for 20 seconds, observing the surrounding virtual environment.

2. **Limited actions**: Wave hands left and right (10 seconds), extend arms (10 seconds), feel the impact of actions on the surrounding environment.

3. **Movement test**: Jump (10 seconds), squat (2 times, holding for 2 seconds), and move left and right (10 seconds) to test the impact of virtual avatar size on dynamic interaction.

After completing the test for each size, participants fill out a questionnaire to evaluate the sense of immersion at that size. The questionnaire includes 7 subjective questions about immersion, interactivity, and spatial depth, optimized based on the Spatial Presence Experience Scale (SPES), using a 5-point Likert scale for quantification

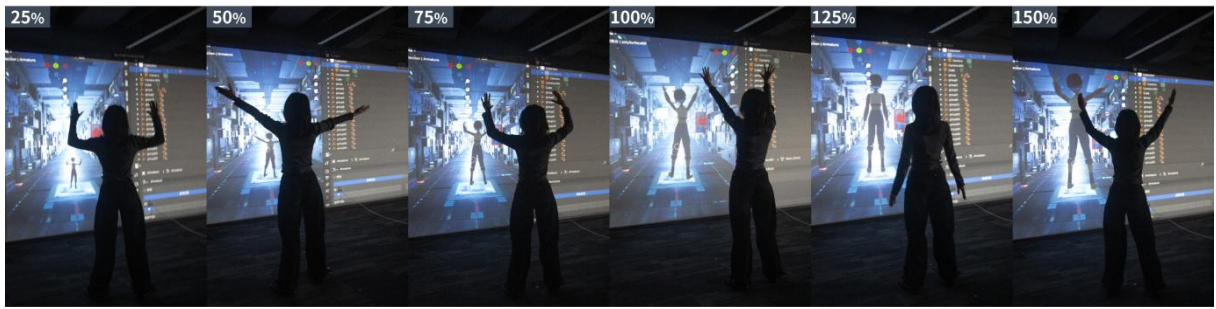

**Figure 4    Experimental Setup for Virtual Avatar Size Preference**

*4.2.4 Experimental Procedure* The participants in Experiment 2 are the same as those in Experiment 1, in order to minimize the impact of sample variation on the experimental results. Detailed information can be found in section 4.1.

*4.2.5 Result* In this experiment, we evaluated the impact of six different virtual avatar sizes (25%, 50%, 75%, 100%, 125%, and 150%) on user immersion (Appendix A Table 2). The 30 participants rated their sense of immersion for each size based on a 5-point Likert scale.

Experimental results show that in the 75%-100% range, immersion scores were significantly higher than other intervals, with higher median and upper quartile scores, indicating a more consistent immersive experience in this range. In cases where sizes deviated significantly (below 50% or above 125%), immersion scores decreased markedly, validating Hypothesis 2 that, as the virtual avatar size deviates, the sense of immersion decreases (Figure 5).

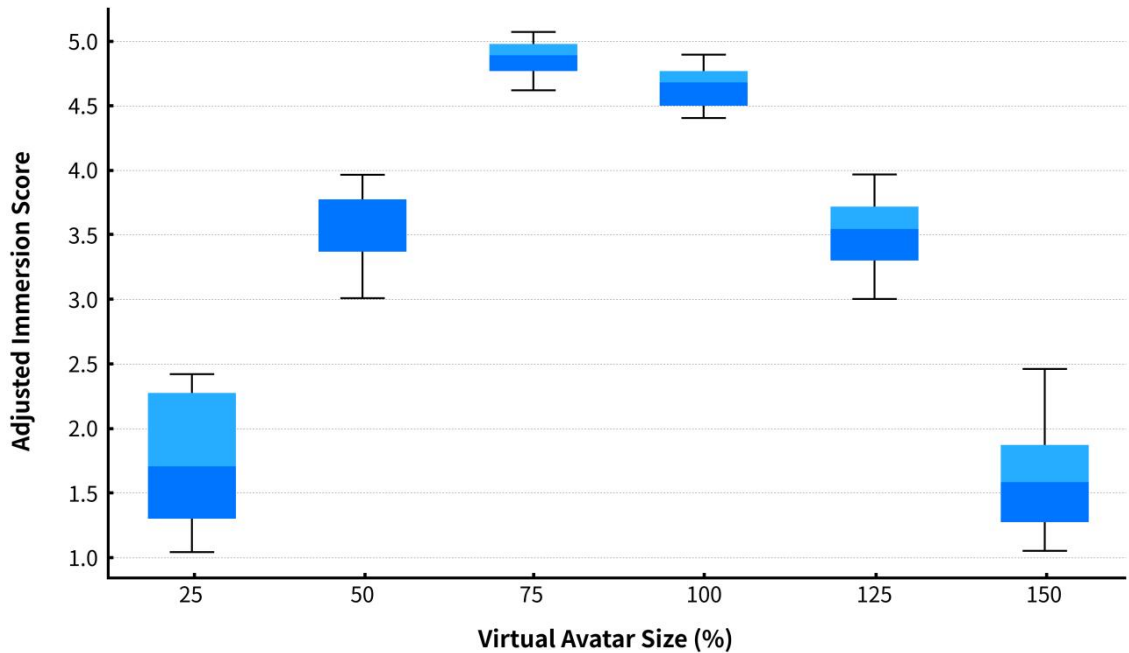

**Figure 5** Boxplot of adjusted immersion scores by virtual avatar size for 30 participants.

To further explore specific differences between sizes, we used Tukey HSD post-hoc tests for pairwise comparisons between groups (Table 1, Figure 6). The results show:

Although we hypothesized that a 1:1 mapping ratio (i.e., 100%) would provide the most natural immersive experience as it directly simulates the real-world space, the experimental results show that participants preferred a ratio of around 130%. This might be because a slightly exaggerated spatial ratio enhanced the visual impact and sense of immersion in the environment. In the 135%-140% range, the frequency of participant choices decreased by about 40% compared to the 130% range, indicating that most participants reported a slight sense of distortion at larger ratios. Especially beyond 140%, the frequency of choices further decreased significantly, supporting Hypothesis 2 that as the ratio increases, the sense of immersion might increase to some extent, but beyond a certain threshold, a sense of distortion begins to emerge and affect the experience.

1. 25% vs. 50%, 75%, 100%, 125%: Significant mean differences ($p < 0.05$), especially between 75%-125% sizes, supporting the hypothesis that sizes closer to the user's actual height can enhance immersion.

2. 25% vs. 150%: No significant difference ($p = 0.79$), indicating that immersion scores at extreme sizes do not differ significantly, but the overall trend suggests lower immersion at these sizes.

3. 75% vs. 100%: No significant difference (p = 0.08), indicating similar immersion performance for these two sizes.

**Table 1  Results for beta-coefficient parameter differences for small targets.**

| Group 1 | Group 2 | Mean Difference | p-value | Lower Bound | Upper Bound | Significant Difference |
|---|---|---|---|---|---|---|
| 25 | 50 | 1.805 | 0 | 1.5686 | 2.0414 | TRUE |
| 25 | 75 | 3.1093 | 0 | 2.8729 | 3.3457 | TRUE |
| 25 | 100 | 2.8863 | 0 | 2.6499 | 3.1227 | TRUE |
| 25 | 125 | 1.7507 | 0 | 1.5143 | 1.9871 | TRUE |
| 25 | 150 | -0.1057 | 0.7912 | -0.3421 | 0.1307 | FALSE |
| 50 | 75 | 1.3043 | 0 | 1.0679 | 1.5407 | TRUE |
| 50 | 100 | 1.0813 | 0 | 0.8449 | 1.3177 | TRUE |
| 50 | 125 | -0.0543 | 0.9857 | -0.2907 | 0.1821 | FALSE |
| 50 | 150 | -1.9107 | 0 | -2.1471 | -1.6743 | TRUE |
| 75 | 100 | -0.223 | 0.0767 | -0.4594 | 0.0134 | FALSE |
| 75 | 125 | -1.3587 | 0 | -1.5951 | -1.1223 | TRUE |
| 75 | 150 | -3.215 | 0 | -3.4514 | -2.9786 | TRUE |
| 100 | 125 | -1.1357 | 0 | -1.3721 | -0.8993 | TRUE |
| 100 | 150 | -2.992 | 0 | -3.2284 | -2.7556 | TRUE |
| 125 | 150 | -1.8563 | 0 | -2.0927 | -1.6199 | TRUE |

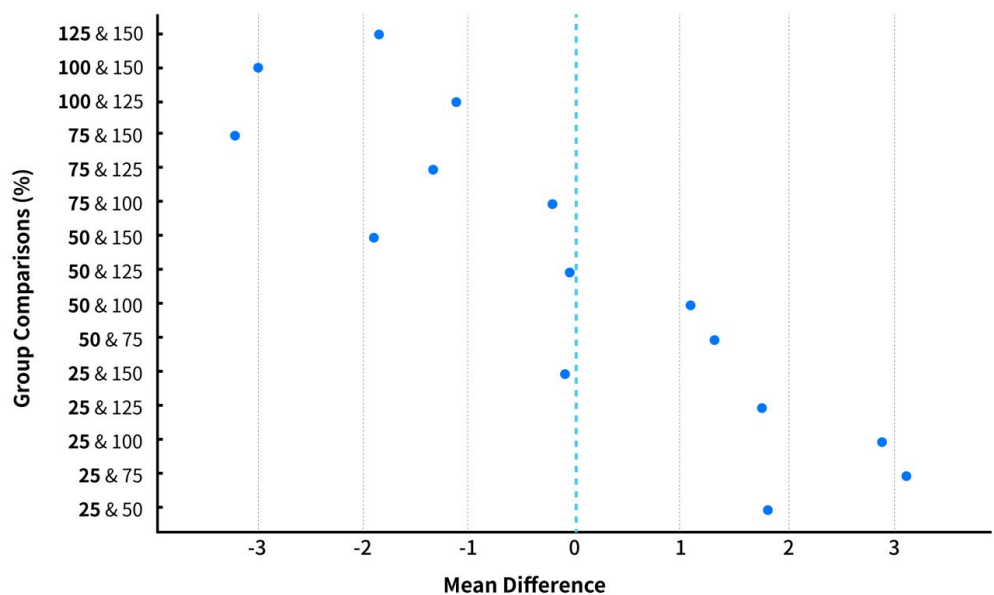

**Figure 6  Tukey HSD test results: Mean differences with 95% CI.**

During the experiment, we also observed differences in perception sensitivity at different ratios: when the virtual avatar size changed from 75% to 100%, 14 participants reported that the change was not noticeable, repeatedly asking operators if the size had been adjusted; while when the size changed from 25% to 75%, 16 participants reported that the perception exceeded expectations. The results suggest that the perception sensitivity of virtual avatar size changes varies by interval, indicating that users' perception of size adjustments is either sluggish or unusually sensitive within certain ratio ranges. These findings suggest that the 75%-100% size range is more conducive to enhancing immersion, while at more extreme ratios, although immersion may decrease, it sometimes provides other types of experiences.

## 5 Exhibition Hybrid Spaces Application

The experimental results have been directly applied to the spatial design and virtual agent configurations in the Intel GTC exhibition hall, located in the World Trade Center, Dongcheng District, Beijing. By integrating the optimized spatial mapping ratios and virtual avatar sizes derived from our research, visitors are able to experience a more authentic fusion of virtual and real elements, showcasing Intel advancements in body-mapping technology and interactive experiences (Figure 7).

Experiments have shown that a spatial mapping ratio of 130% can enhance both immersion and visual impact. Therefore, we set the virtual space dimensions to 3.25 meters (2.5 meters × 1.3) by 2.6 meters (2 meters × 1.3) in the GTC exhibition hall. This slightly exaggerated ratio enhances the sense of "envelopment" in the environment. In determining the appropriate size of the virtual avatar, we considered the average ergonomic dimensions of Chinese males and females (male: 169.7 cm; female: 158.0 cm) to ensure an inclusive design. According to our experimental results, setting the virtual avatar's height to 75% optimizes the user's sense of embodiment and interaction experience with the virtual avatar while avoiding the sense of distortion that can arise from mismatched sizes. Therefore, we set the virtual agent's height to approximately 127 cm (75% of the average male height and 80.55% of the average female height), aligning with the recommended range of 75% to 100% of the user's height and is consistent with previous research recommendations on virtual avatar. This design helps to create a virtual environment that feels natural and comfortable for most users.

These experimental results provide practical guidance for spatial design and virtual avatar settings for immersion in exhibition hybrid spaces. By applying these optimised parameters, we were able to increase user immersion, reduce perceived distortion, and thereby improve the overall quality of the interactive experience.

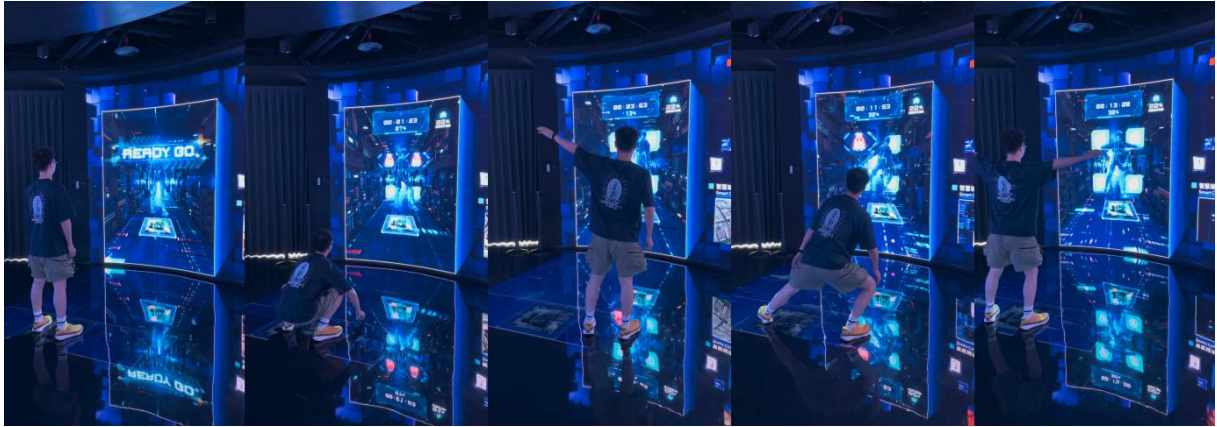

Figure 6　Immersive interactive experience at the Intel GTC exhibition hall

## 6 Discussion

This study aimed to further explore the balance between real and virtual spatial scale consistency to enhance immersion in exhibition hybrid spaces. Through user testing with virtual avatars and virtual spaces at varying scales, we investigated how different dimensions affect the sense of immersion. Overall, participants reported optimal immersion when virtual space was larger than real space, while virtual avatar was smaller than user. This suggests that deliberately adjusting the consistency of scale between real and virtual elements can improve immersion in exhibition hybrid environments.

In terms of the scale of the virtual space, the size of 130% provided participants with greater sense of immersion. 68.3% of participants reported that a virtual space slightly larger than the real space offered a sense of envelopment, and felt "I am truly here." At this scale, users' cognition could be more attuned to the information presented in the virtual environment, fully immersing them and preventing the real environment from interfering with the interactive experience. In terms of virtual avatar size, the results showed that participants experienced greater immersion in the 75%-100% scale range. When interacting in the virtual space, this slightly smaller virtual avatar created a stronger alignment between participants' behaviour and cognition. This alignment reduced the sense of identity distance between users and their virtual avatars, even allowing users to embody the virtual avatar and act in unison, resulting in a deeper sense of immersion or a feeling of "it's me". In addition, participants were able to perceive changes in the size of the virtual avatar between 25% and 75% keenly and accurately. However, perception was relatively weak for scale changes between 75%-100%. The results suggest that the perception sensitivity to changes in the size of the virtual avatar varies by interval. In contrast to the immersive experience, when the virtual avatar was scaled to 25%, participants felt a stronger sense of avatar embodiment and character, making it more suitable for narrative-driven in hybrid space. The combined application of these two scales in hybrid spaces allows users to more easily understand interactive content and tasks, achieving physiological and psychological consistency between real and virtual environments. This enables users to establish a deeper connection with the virtual world, thus meeting their requirements for an immersive experience.

Through analysis of the experimental data, we also found that virtual space perspective similarly affects user immersion. Furthermore, it is expected that a formula for the optimal experience in immersive spaces can be derived, allowing the calculation of ideal virtual space and avatar scales based on real-world measurements once a larger dataset is obtained. These findings will be important directions for further research.

However, there are some limitations to this study. Although Intel 3D Athlete Tracking (3DAT) is a stable motion capture technology, its high sensitivity can amplify users' motion responses, making actions more prominent on the screen and potentially distracting participants' attention. Additionally, despite using real-time motion capture technology, there was still some delay in the transmission of motion signals during the experiments. Finally, the age range of participants was relatively homogeneous; future studies will include a broader age range to increase the generalisability and reliability of the findings. Future research will aim to address these limitations and further optimise the immersive space interaction experience.

## 7 Conclusion

This study explored how intentional adjustments to the spatial scales of virtual environments and avatars can enhance user immersion in exhibition hybrid spaces. By strategically manipulating these scales, we found that aligning users' behaviour and cognition facilitates a deeper connection with the virtual world without relying on wearable devices. This approach reduces cognitive dissonance and provides practical guidelines for designing immersive hybrid spaces.

While our research focused on exhibition settings, the findings may be applicable to other domains involving the integration of the virtual and real worlds, such as education, training, and entertainment. Future work will address current limitations—such as motion capture sensitivity and participant diversity—by refining the technology and including a more diverse sample. By improving the consistency between real and virtual environments, this study provides valuable insights for practitioners aiming to enhance immersion in various interactive contexts.

## Acknowledgments

The 3DAT technology used in this study is supported by Intel Corporation and Beijing Xuanmi Technology Co., Ltd.

## Funding

This research was supported by the "Dual High" Project of Tsinghua Humanity Development (No. 2023TSG08103).

## Ethical Approval and Informed Consent

This study involved human participants. According to the regulations of Tsinghua University, formal ethical approval was not required for this type of research. All participants were informed about the purpose of the study and provided their informed consent prior to participation.